\documentclass[conference]{IEEEtran}
\IEEEoverridecommandlockouts
\usepackage{cite}
\usepackage{amsmath,amssymb,amsfonts}
\usepackage{algorithmic}
\usepackage{graphicx}
\usepackage{textcomp}
\usepackage{xcolor}
\usepackage{booktabs}
\usepackage{multirow}
\def\BibTeX{{\rm B\kern-.05em{\sc i\kern-.025em b}\kern-.08em
    T\kern-.1667em\lower.7ex\hbox{E}\kern-.125emX}}
\begin{document}

\title{Multi-scale Transformer-based Network for Emotion Recognition from Multi Physiological Signals
\thanks{This work was supported by the National Research Foundation of Korea (NRF) grant funded by the Korea government (MSIT) (NRF-2020R1A4A1019191). This work was also supported by Institute of Information \& communications Technology Planning \& Evaluation (IITP) grant funded by the Korea government(MSIT) (No.2021-0-02068, Artificial Intelligence Innovation Hub). (\textit{Corresponding author: Soo-Hyung Kim.})
}

}

\author{\IEEEauthorblockN{Tu Vu\textsuperscript{\textdagger}, Van Thong Huynh\textsuperscript{\textdagger}, Soo-Hyung Kim}
\IEEEauthorblockA{\textit{Department of AI Convergence} \\
\textit{Chonnam National University}\\
Gwangju, South Korea \\
\{tu369,vthuynh,shkim\}@jnu.ac.kr}
}

\maketitle
\begingroup\renewcommand\thefootnote{\textdagger}
\footnotetext{Equal contribution.}
\endgroup

\begin{abstract}
This paper presents an efficient Multi-scale Transformer-based approach for the task of Emotion recognition from Physiological data, which has gained widespread attention in the research community due to the vast amount of information that can be extracted from these signals using modern sensors and machine learning techniques. Our approach involves applying a Multi-modal technique combined with scaling data to establish the relationship between internal body signals and human emotions. Additionally, we utilize Transformer and Gaussian Transformation techniques to improve signal encoding effectiveness and overall performance. Our model achieves decent results on the CASE dataset of the EPiC competition, with an RMSE score of 1.45.
\end{abstract}


\section{Introduction}

Recognizing emotions is a fundamental aspect of human communication, and the ability to accurately detect emotional states has significant impacts on a range of applications, from healthcare to human-computer interaction. Emotions are often reflected in physiological signals~\cite{shu2018review}, facial~\cite{li2020deep}, and speech~\cite{schuller2018speech}. Recently, the use of physiological signals for affective computing has gained considerable attention due to its potential to provide objective measures of emotional states in real-time~\cite{saganowski2022emotion}.

Recently, there has been a growing interest in developing machine learning algorithms for affective computing using physiological signals~\cite{dominguez2020machine,vazquez2022emotion,santamaria2018using,harper2020bayesian, algarni2022deep}. These algorithms can be used to classify emotional states, predict changes in emotional states over time, or identify the specific features of physiological signals that are most informative for detecting emotional states. There has also been interested in developing wearable sensors that can capture physiological signals in real-world settings, such as in the workplace or in social situations~\cite{saganowski2021system}.

The use of end-to-end deep learning architectures for physiological signals has the potential to simplify the development and deployment of an emotion recognition system \cite{saganowski2022emotion}. By eliminating the need for preprocessing steps, these architectures can reduce the complexity and time required for system development, as well as improve the scalability and accuracy of the system. Moreover, end-to-end architectures can enable the development of systems that can process multiple physiological signals simultaneously, such as heart rate, respiration, and electrodermal activity, which can provide more comprehensive and accurate measures of emotional states.

Despite the potential benefits of end-to-end deep learning architectures for affective computing, there are still challenges that need to be addressed. One challenge is to develop architectures that can handle noisy and non-stationary physiological signals, which can be affected by movement artifacts, signal drift, and other sources of noise. Another challenge is to ensure that the learned features are interpretable and meaningful, which can help improve the transparency and explainability of the system.

In this paper, we propose an end-to-end Multi-scale architecture for continuous emotion regression with physiological signals. We evaluate the performance of the proposed architecture using EPiC 2023 competition benchmark, which contains data collected from experiments conducted in a laboratory setting. 

\section{Related Works}
\subsection{Continous emotion recognition from multimodal physiological signal}

The utilization of physiological signals has been widely acknowledged as one of the most reliable data forms for affective science and affective computing. Despite the fact that individuals are capable of manipulating their physical signals such as facial expressions or speech, consciously controlling their internal state is quite a daunting task. Therefore, analysis of signals from the human body represents a dependable and robust approach for fully recognizing and comprehending an individual's emotional state \cite{shu2018review, ahmad2022survey}. This reliability factor is especially crucial in medical applications, such as mental health treatment or mental illness diagnosis.

Recognizing affect from physiological data remains a significant challenge, not only during the data acquisition process but also in terms of emotion assessment. Laboratory-based research dominates the field of affective science due to the control it affords over experimental variables. Researchers can carefully select and prepare emotional stimuli, and employ various sensor devices to trace and record a subject's emotional state with minimal unexpected event, interference \cite{saganowski2022emotion}.  However, most of these studies rely on discrete, indirect methods such as quizzes, surveys, or discrete emotion categories for emotion assessment, which overlook the time-varying nature of human emotional experience. Sharma et al. \cite{sharma2019dataset} introduced a novel Joystick-based Emotion Reporting Interface (JERI) to overcome a limitation in emotion assessment. JERI enables the simultaneous annotation of valence and arousal, allowing for moment-to-moment emotion assessment. The Continuously Annotated Signals of Emotion (CASE) dataset, acquired using JERI, provides additional information to researchers for identifying the timing of emotional triggers. Leveraging this dataset, the Emotion Physiology and Experience Collaboration (EPiC) Workshop organized a competition to model moment-to-moment, continuous ratings of valence and arousal from physiological signals. This competition serves as a valuable platform for evaluating the precision and temporal dynamics of emotion recognition models. Accordingly, we adopt the EPiC competition benchmark as our primary benchmark for evaluating our proposed emotion recognition model.

Besides, it is claimed that a single physiological signal are relatively difficult to precisely reflect human's emotional changes. Hence, recently, there have been many researches focusing on detecting human's emotion on Multimodal Physiological Signal. There are many type of physiological signal used in these studies. While some study record heart related signals such as electrocardiographic (ECG) \cite{hu2018scai, rattanyu2010emotion, rattanyu2011emotion}, blood volume pulse (BVP) \cite{zhao2018emotionsense, ragot2018emotion}, others use electrical activity of the brain (Electroencephalogram/EEG) \cite{martens2020sensor, nakisa2018long} or muscle's electrical reaction (Electromyogram/EMG) \cite{schmidt2019multi, romeo2019multiple}. Furthermore, somes even employ 
skin temperature (SKT) \cite{romeo2019multiple}, skin sweat glands (EDA)\cite{schmidt2019multi,martens2020sensor} and the depth and rate of breathing (Respiratory/RSP)\cite{schmidt2019multi}. Based on signal modal provide by EPiC competition, our method utilized 8 different type of signals including ECG, BVP, EMG\_CORU (corrugator supercilii), EMG\_TRAP (trapezius), EMG\_ZYGO (zygomaticus major), GRS, RSP and SKT.

\subsection{Transformer based method for in multimodal emotion recognition from physiological signal}
Similar to other emotion recognition problems that involve physical signals, Affective Computing in physiological data has witnessed extensive adoption of Machine Learning techniques, particularly Deep Learning methodologies. In their work, Dominguez et al \cite{dominguez2020machine} employed various conventional Machine Learning techniques, including Gaussian naive Bayes, k-Nearest Neighbours, and Support Vector Machines, for Valence-Arousal estimation. However, these approaches are heavily dependent on the quality of handcrafted feature selection and feature extraction processes. To overcome this challenge, other studies \cite{vazquez2022emotion, santamaria2018using, harper2020bayesian} proposed the use of Deep Learning techniques for an end-to-end approach, where the model learns to extract features automatically without the need for pre-designed feature descriptors.

With the advancement of Deep Learning, various state-of-the-art techniques have been employed for Affective Analysis from Physiological signals. Santamaria et al. \cite{santamaria2018using} used Convolutional Neural Networks (CNNs) with 1D Convolution layers for Emotion Detection, while Harper et al. \cite{harper2020bayesian} combined CNNs with frequently used Recurrent Neural Networks (RNNs) for Emotion Recognition from ECG signals. Since their introduction in 2016, Transformers \cite{vaswani2017attention} have emerged as the preferred models in the field of Deep Learning. Their robust performance in Natural Language Processing, a type of data that shares some similar characteristics with time-series data, has demonstrated the potential of Transformers when applied to time-series signals. As a result, recent research in the Time-Series domain has utilized Transformers as the core module in their model architecture \cite{li2019enhancing, wu2020deep, li2021dct}. For physiological signals, some studies have proposed using Transformers and their variants for detecting emotions \cite{vazquez2022transformer, vazquez2022emotion, wu2023transformer, yang2022mobile}. In the works of Vazquez et al. \cite{vazquez2022transformer, vazquez2022emotion}, they focused on applying pretrained Transformers for Multimodal signal processing. However, this is still a very basic application of Transformer modules. Wu et al. \cite{wu2023transformer} and Yang et al. \cite{yang2022mobile} proposed using more advanced techniques of Transformer-based models, which are self-supervised and Convolution-augmented Transformer for Single and Multimodal signal processing. Although these studies have demonstrated the effectiveness of Transformers for physiological signals, they often feed the model with fixed original size signals, which may lead to the loss of global feature information. To address this issue, we propose a new Multi-scale Transformer-based architecture for Multimodal Emotion Recognition.

\section{Proposed approach}
\begin{figure*}
    \centering
    \includegraphics[width=.9\linewidth]{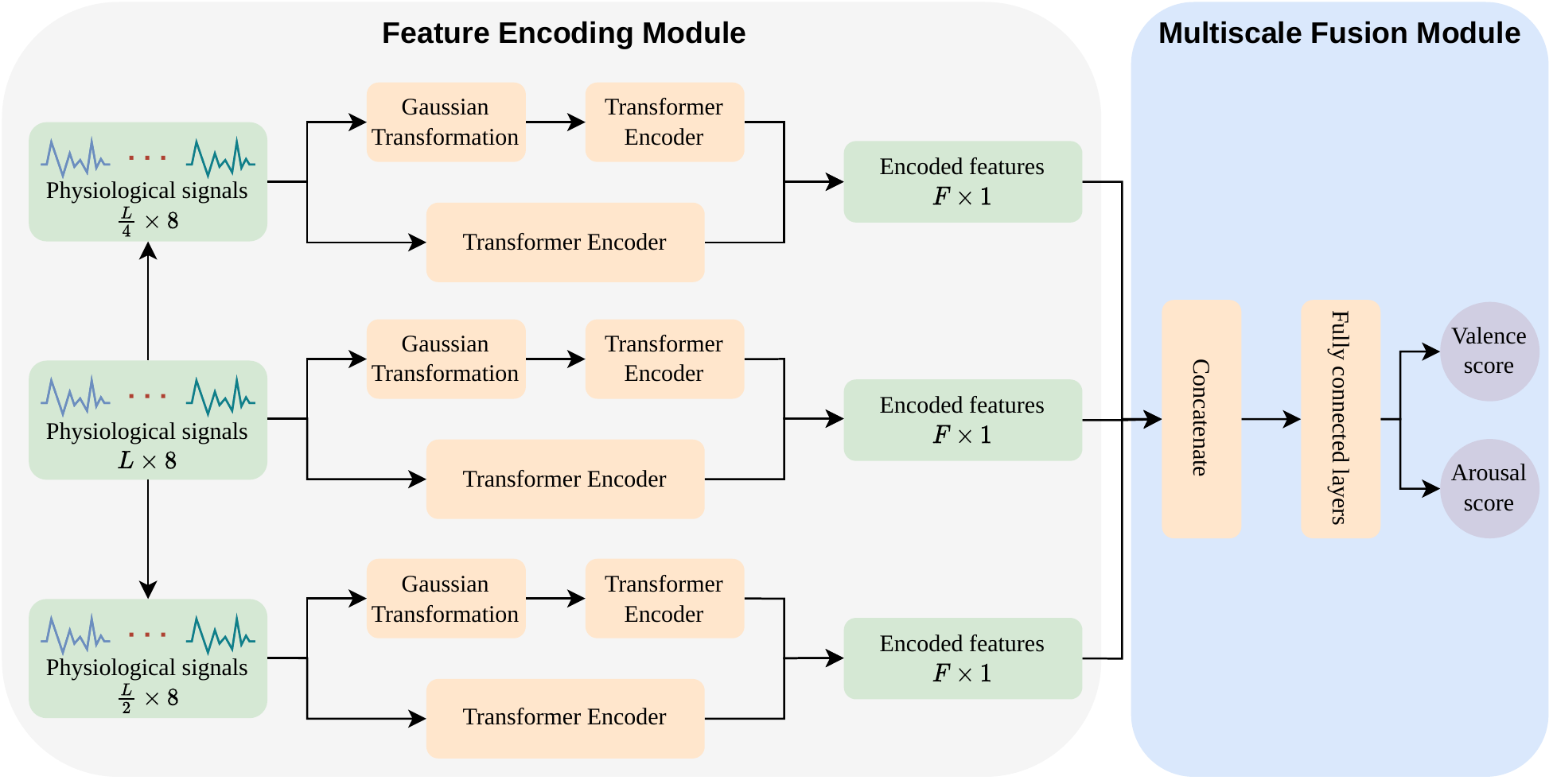}
    \caption{An overview of our proposed architecture.}
    \label{fig:epicArch}
\end{figure*}

\subsection{Problem definition}
The  Emotion Recognition in Multimodal Physiological Signal problem takes as input 8 physiological signals, namely ECG, BVP, EMG\_CORU, EMG\_TRAP, EMG\_ZYGO, GRS, RSP and SKT, extracted from human subjects during emotion-inducing stimuli. This denoted as 8 sequence with L length. In Affective computing field, the Emotion Recognition problem objective varies relying on indicated the Emotional Models. In the scope of this study, following the use of SAM (Self-Assessment Manikin) \cite{bradley1994measuring} model of the CASE dataset, the problem objective is Valence-Arousal (V-A) estimated value. The V-A score consists of two continuous floating-point numbers ranging from 0.5 to 9.5. A value of 0.5 denotes the most negative valence or lowest arousal, 5 indicates neutral valence or arousal, and 9.5 indicates the most positive valence or highest arousal.

\subsection{Methodology}
We construct a new Multi-scale architecture for Valence-Arousal estimation from 8 Physiological signals. Our architecture comprises of two core modules: Feature encoding module and Multi-scale Fusion module. Overall architecture is showed in Figure \ref{fig:epicArch}.

\subsubsection{Feature encoding}
To enable the feature encoding module to extract global features for the estimator and eliminate noise and interference information from the input, we employ 1-Dimensional Average Pooling to scale the 8 input signals into three different lengths: $L$, $L/2$, and $L/4$. This process helps to improve the model's ability to extract useful information and eliminate unwanted noise and interference. 

Then, we simultaneously apply two kind of feature encoder which is Gaussian Transform \cite{rahimi2007random} and Transformer Encoder \cite{vaswani2017attention}. The Transformer Encoder block is used Multi-Headed Self-Attention as it core mechanism. 
Given an input sequential signal $S \in R^{L \times C}$, where $L$ represents the length of the signal sequence and $C = 8$ is the number of channels (signal modalities), we apply a Positional Encoding and Embedding layer to convert the raw input into a sequence of tokens. Subsequently, the tokens are fed into Transformer layers consisting of Multi-Headed Self-Attention (MSA) \cite{vaswani2017attention}, Layer Normalization (LN), and Multi-Layer Perceptron (MLP) blocks. Each element is formalized in the following equations:

\begin{equation}
  y^i = MSA(LN(x^i)) + x^i
  \label{eq:important}
\end{equation}
\begin{equation}
  x^{l+i} = MLP(LN(y^i)) + y^i
  \label{eq:important}
\end{equation}
Here, $i$ represents the index of the token, and $x^i$ denotes the generated feature's token. It is worth noting that since the Multi-Headed Self-Attention mechanism allows multiple sequences to be processed in parallel, all 8 signal channels are fed into the Transformer Encoder at once.

The Gaussian Transform  \cite{rahimi2007random} is traditionally employed to kernelize linear models by nonlinearly transforming input features via a single layer and subsequently training a linear model on top of the transformed features. However, in the context of deep learning architectures, random features can also be leveraged, given their ability to perform dimensionality reduction or approximate certain functions via random projections. As a non-parametric technique, this transformation maps input data to a more compressed representation that exclude noise information while still enabling the computationally efficient processing. Such a technique may serve as a valuable supplement to Transformer Encoder architectures, compensating for any missing information.

\subsubsection{Multi-scale Fusion}
From the feature extracted from the Feature Encoder Module at different scale, we fuse them by using Concatenation operation. The concatenated features are then fed through a series of Fully-connected layers (FCN) for the estimation of the 2 Valence and Arousal scores. The Rectified Linear Unit (ReLU) activation function is chosen for its ability to introduce non-linearity into the model, thus contributing to the accuracy of the score estimation. The effectiveness of this approach lies in its ability to efficiently estimate the desired scores, while maintaining a simple and straightforward architecture.

\section{Experimental and Results}
\subsection{Dataset}
Following the Emotion Physiology and Experience Collaboration (EPiC) Workshop, we evaluate our model using the Continuously Annotated Signals of Emotion (CASE) dataset  \cite{sharma2019dataset}.
 The dataset contains data from several physiological sensors and continuous annotations of emotion. This data was acquired from 30 subjects while they watched several video-stimuli and simultaneously reported their emotional experience using JERI. The devices used include sensors for Electrocardiograph (ECG), Blood Volume Pulse (BVP), Galvanic Skin Response (GSR), Respiration (RSP), Skin Temperature (SKT), and Electromyography (EMG). These sensors returns 8 type of Physiological signals: ECG, BVP, EMG\_CORU, EMG\_TRAP, EMG\_ZYGO, GRS, RSP and SKT. The emotional stimuli consisted of 11 videos, ranging in duration from 120 to 197 seconds. The annotation and physiological data were collected at a sampling rate of 20 Hz and 1000 Hz, respectively. The initial Valence-Arousal score range was set at [-26225, 26225].

In EPiC Competition, the organizers restructure the data and scale the annotation value into [0.5, 9.5] range. The reconstructed dataset has 4 scenarios for 4 different evaluation approach: Across-time scenario, Across-subject scenario, Across-elicitor scenario and Across-version scenario. 
\begin{itemize}
  \item Across-time scenario: Each sample represents a single person watching a single video, and the training and test sets are divided based on time. Specifically, the earlier parts of the video are used for training, while the later parts are reserved for testing. 
  \item Across-subject scenario: Participants are randomly assigned to groups, and all samples from a given group belong to either the train or test set depending on the fold.
  \item Across-elicitor scenario: Each subject has two samples (videos) per quadrant in the arousal-valence space. For each fold, both samples related to a given quadrant are excluded, resulting in four folds, with one quadrant excluded in each fold.
  \item Across-version scenario: Each subject has two samples per quadrant in the arousal-valence space. In this scenario, one sample is used for training the model, and the other sample is used for testing, resulting in two folds.
\end{itemize}

\subsection{Experiments setup}
Our networks were implemented using the Tensorflow framework. We trained our models using the AdamW optimizer \cite{loshchilov2017decoupled} with a learning rate of 0.001 and the Cosine annealing warm restarts scheduler \cite{loshchilov2016sgdr} over 10 epochs. The MSE loss function was used to optimize the network. RMSE is the metrics for evaluation stage. The sequence length was set to 2048. We utilized 4 Transformer layers for the Transformer Encoder, with each Attention module containing 4 heads. The hidden dimension of the Transformer was set to 1024. All training and testing processes were conducted on a GTX 3090 GPU.
\begin{table}[]
  \centering
    \caption{Result on the Test Data on Scenarios-level.}
  \resizebox{\columnwidth}{!}{
\begin{tabular}{lcccc} \toprule
 \multirow{2}{7em}{Scenarios type} & \multicolumn{2}{c}{Arousal} & \multicolumn{2}{c}{Valence} \\ \cmidrule(lr){2-3} \cmidrule(lr){4-5}
           & RMSE         & STD           & RMSE         & STD          \\ \midrule
Across-time scenario     & 1.503        & 1.083         & 1.639        & 1.2246       \\
Across-subject scenario  & 1.336        & 0.2276        & 1.345        & 0.0576       \\
Across-elicitor scenario & 1.509        & 0.6771        & 1.514        & 0.6054       \\
Across-version scenario  & 1.369        & 0.2525        & 1.352        & 0.1919       \\\bottomrule
\end{tabular}
\label{tab:Result1}
} 
\end{table}

\subsection{Results}

Table \ref{tab:Result1} presents the results of our model on the test set in terms of scenario-level evaluation. Overall, the final RMSE score for both Valence and Arousal estimation task we gain is 1.45. Our model achieved the best performance in the Across-subject scenario, with an Arousal score of 1.336 and a Valence score of 1.345, with relatively low standard deviation of 0.2276 and 0.0576 for Arousal and Valence, respectively. These results suggest that our model can effectively generalize to new subjects and accurately capture the emotion change  after fully viewing the entire video-viewing process. Meanwhile, the relatively low performance in the Across-elicitor scenario, with scores of 1.509 and 1.514 in Arousal and Valence, respectively, suggests that our model did not perform well in inferring emotional states that were not seen during training, given the previously learned specific emotional states.

\section{Conclusion}
This paper proposes a new Multi-scale architecture for Multimodal Emotion Recognition from Physiological signals. Our approach involves encoding the signal with Transformer Encoder at multiple scales to capture both global and local features and obtain more informative representations. Our method achieved decent results on the test data of the EPiC Competition.

\section*{Ethical Impact Statement}

Our study leveraged deep learning methods for emotion estimation from physiological signals has ethical implications regarding potential bias, generalizability. The data used for model training was collected from a specific population, introducing the possibility of bias and limiting the generalizability of the models to other populations. Physiological signals can vary among individuals, affecting the accuracy of emotion estimation and further reducing generalizability. We acknowledge these limitations and recommend future research to address these issues and develop ethical guidelines to ensure appropriate use of the models.

\bibliographystyle{ieeetr}
\bibliography{refs}

\end{document}